\documentclass[journal]{IEEEtran}
\IEEEoverridecommandlockouts
 
\usepackage{booktabs}
\usepackage{bbm}
\usepackage{amsmath}
\usepackage{mathtools}
\usepackage{amsfonts}
\usepackage{float}
\usepackage{cite}
\usepackage{caption}
\usepackage{subcaption}
\usepackage{multirow}
\usepackage{color,soul}
\usepackage{multirow}
\usepackage[flushleft]{threeparttable}
\usepackage[ruled,linesnumbered]{algorithm2e}  
\usepackage{array} 
\usepackage{booktabs}

\SetKwInput{KwData}{\textbf{Input}}

\let\oldnl\nl
\newcommand{\nonl}{\renewcommand{\nl}{\let\nl\oldnl}}
\begin{document}
\title{Deep Task-Based Beamforming and Channel Data Augmentations for Enhanced Ultrasound Imaging}
\author{Ariel Amar,  
Ahuva Grubstein,  
Eli Atar,
Keren Peri-Hanania,
Nimrod Glazer,  
Ronnie Rosen,  
Shlomi Savariego, 

and Yonina C. Eldar, \IEEEmembership{Fellow, IEEE}  
\thanks{
This research is supported by an Amazon fellowship.

A. Amar, K. Peri-Hanania, N. Glazer, R. Rosen, S. Savariego, and Y. C. Eldar are with the Faculty of Math and CS, Weizmann Institute of Science, Rehovot, Israel, email:\{ariel.amar, keren.peri, nimrod.glazer, ronnie.rosen, shlomi.savariego, yonina.eldar\}@weizmann.ac.il. A. Grubstein, E. Atar are with Rabin Medical Center, Petah Tikva, Israel, and Faculty of Medical and Health Sciences, Tel Aviv University, Tel Aviv, Israel, email:\{ahuvag, elia\}@clalit.org.il.}  

\vspace{-0.5cm}

}

\maketitle

\begin{abstract} This paper introduces a deep learning (DL)-based framework for task-based ultrasound (US) beamforming, aiming to enhance clinical outcomes by integrating specific clinical tasks directly into the beamforming process. Task-based beamforming optimizes the beamformer not only for image quality but also for performance on a particular clinical task, such as lesion classification. The proposed framework explores two approaches: (1) a Joint Beamformer and Classifier (JBC) that classifies the US images generated by the beamformer to provide feedback for image quality improvement; and (2) a Channel Data Classifier Beamformer (CDCB) that incorporates classification directly at the channel data representation within the beamformer's bottleneck layer. Additionally, we introduce channel data augmentations to address challenges posed by noisy and limited in-vivo data. Numerical evaluations demonstrate that training with channel data augmentations significantly improves image quality. The proposed methods were evaluated against conventional Delay-and-Sum (DAS) and Minimum Variance (MV) beamforming techniques, demonstrating superior performance in terms of both image contrast and clinical relevance. Among all methods, the CDCB approach achieves the best results, outperforming others in terms of image quality and clinical relevance. These approaches exhibit significant potential for improving clinical relevance and image quality in ultrasound imaging. \end{abstract}

\section{Introduction}
\label{section:Introduction}

A standard ultrasound probe typically consists of multiple piezoelectric elements, also known as channels. These elements are dual-functional; they emit pressure waves and capture the echoes reflected back. Image formation from these echoes involves a process called beamforming, which typically unfolds in three stages:

\begin{enumerate}
    \item \textbf{Time of Flight (ToF) Correction} - This stage involves converting the signal samples from the time domain to the space domain to align the timing of the received echoes accurately.
    \item \textbf{Apodization} - This involves applying a weighted sum across the channel dimension to mitigate the sidelobes of the ultrasound beam, enhancing image clarity.
    \item \textbf{Imaging} - This final stage incorporates several post-processing steps such as envelope detection, logarithmic compression, and interpolation to transition the pre-processed ultrasound image to the desired image grid.
\end{enumerate}

Beamformers mainly differ in their approach to the apodization stage. Techniques range from non-adaptive methods like DAS to adaptive methods such as Weiner Beamforming and Coherence Factor (CF) \cite{CoherenceFactor2010}, iterative Maximum-a-Posteriori (iMAP)\cite{iMAP2019} and MV \cite{MV}.

Currently, DAS is predominantly used in clinical settings due to its simplicity and capability for real-time image reconstruction. However, although the MV approach is regarded as superior in terms of image quality, its high computational demands make it impractical for real-time clinical use. To bridge the gap between the real-time capabilities of DAS and the superior image quality of MV, recent developments have introduced various DL techniques into the beamforming process. Recent reviews of DL applications in ultrasound imaging have highlighted the significant advances in balancing computational efficiency with high-quality image reconstruction, providing new frameworks that address the limitations of traditional methods while encouraging further innovation in DL-based beamforming \cite{yonina2020review, song2024survey, hyun2021deep}. Several studies have explored specific DL-based solutions for ultrasound beamforming, demonstrating improvements in both speed and image quality across a variety of setups and methods \cite{yoon2018efficient, luijten2020adaptive, khan2019deep, khan2020adaptive, khan2021switchable, li2020beamforming, wang2020conditional, chennakeshava2020high, pilikos2021single, tang2021plane, tierney2021training, simson2019end, lu2022improving, khan2022phase, mathews2021towards, yoon2018deep, luchies2018deep, mamistvalov2021deep, mamistvalov2022deep}.

Recent advances in ultrasound beamforming have explored various auxiliary tasks aimed at enhancing image quality and diagnostic value. For example, research has been directed towards simultaneous ultrasound beamforming and segmentation, enhancing clinical assessment \cite{Nair2020-hk, segmenationAndBeamforming2021}. Moreover, efforts have been made to address inherent ultrasound imaging challenges such as speckle noise and clutter. Techniques focused on speckle reduction and noise suppression, clutter removal \cite{Mor2020-uc,NoiseSupression2021, clutter2020yonina} aim to improve image clarity and contrast, thereby facilitating better diagnosis and measurement accuracy. 

Despite significant advancements in DL-based US beamforming, many existing methods do not tailor the beamforming process to specific diagnostic tasks. In this work, we introduce task-based beamforming, an approach that integrates clinical tasks directly into the beamforming process to enhance both image quality and diagnostic relevance. We focus on breast lesion classification as a representative clinical task to demonstrate our methodology.

We explore two strategies for incorporating clinical task information into deep learning-based beamformers. The first involves multitask learning, where the beamformer is trained alongside an image classifier to optimize both beamforming and classification simultaneously. The second strategy embeds a classification head directly into the beamforming model at the channel data level, allowing the beamformer to receive direct feedback from the clinical task during training. By integrating clinical tasks into the beamforming process, we aim to produce images that are not only of higher quality but also more informative for diagnostic purposes.

Training these models with in-vivo data presents challenges such as the absence of ground truth (GT) images, high noise levels, and limited data availability. To address the lack of GT images, we use MV beamformed images as surrogate labels, following previous studies \cite{luijten2020adaptive,simson2019endtoend}. To mitigate noise and augment the limited dataset, we introduce channel data augmentations, which are corruptions applied to the channel data to help the model become more robust to data loss and noise. These augmentations simulate various physical effects that occur during US scans, enhancing the model's ability to handle real-world scenarios.

Our channel data augmentations include adding speckle noise, applying subsampling and masking, and performing augmentations on the spectrograms of the channel data in both time and frequency domains. By introducing speckle noise to the channel data, we simulate the physical noise commonly encountered during ultrasound scans, allowing the model to better manage the inherent noise in US imaging. Subsampling and masking randomly omit or zero out sections of the channel data, encouraging the model to learn from incomplete or corrupted inputs, thus improving its robustness to data loss. Additionally, we introduce an augmentation method called \textit{Channel Data SpecAugment}, inspired by techniques in speech processing. This method applies random masking and stretching to the spectrograms of the channel data in both time and frequency domains, helping the model become more resilient to temporal and spectral distortions.

Our numerical results demonstrate that training deep learning-based task-based beamformers with these channel data augmentations significantly improves image quality. Both task-based beamformers outperformed traditional methods like DAS and MV, as well as a baseline deep learning-based beamformer trained without a clinical task. Notably, the CDCB beamformer, which integrates the classification task at the channel data level, achieved superior performance. This suggests that utilizing channel data information for the clinical task is more effective than relying solely on clinical information extracted from the beamformed US image.

The remainder of this paper is organized as follows: Section \ref{sec:task_based_beamforming} discusses task-based beamforming and elaborates on the different approaches, JBC and CDCB. Section \ref{sec:cd_aug} describes the channel data augmentation techniques we developed to improve the robustness of our models. Section \ref{sec:res} presents the experimental setup, and the performance of our methods, comparing them to conventional DAS and MV beamforming techniques. Finally, the conclusions are provided in Section \ref{sec:conc}.

\section{Task-Based Beamforming}
\label{sec:task_based_beamforming}

This section presents our DL-based framework for task-based ultrasound beamforming. We investigate approaches to integrate clinical tasks into DL-based ultrasound beamformers.
Here, we focus on the task of breast lesion classification. However, the framework we propose is designed to be versatile and applicable to a wide range of tasks beyond classification, such as regression, detection, and segmentation.

\subsection{Methods for Task-based Beamforming}
\label{sec:Task-based-Beamforming}

\begin{figure*}[ht]
    \centering
    \includegraphics[width=0.8\textwidth]{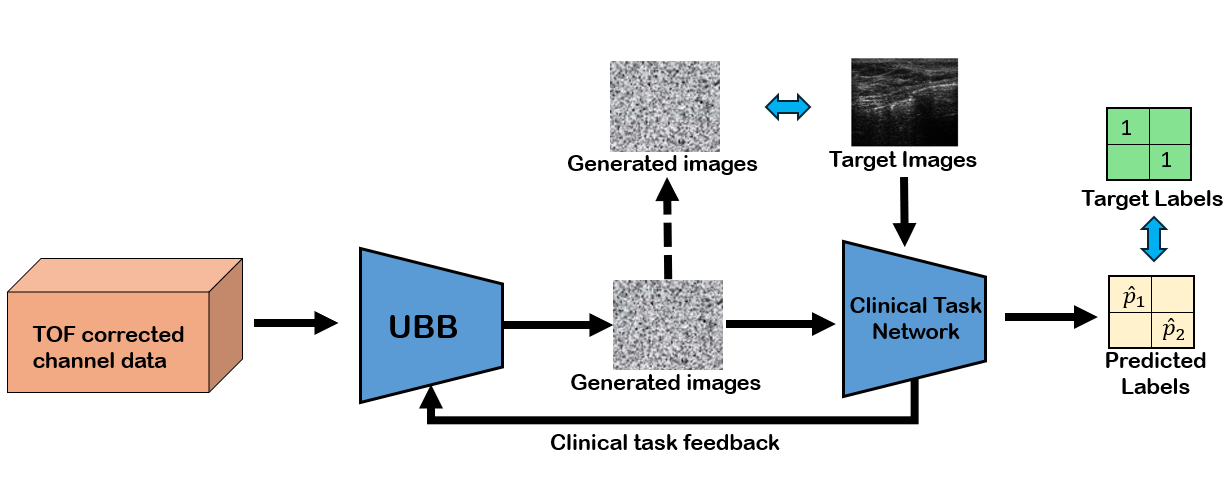}
    \caption{In the JBC approach, the UBB model and the Clinical Task Network are trained simultaneously with separate optimizers. The UBB generates ultrasound images from TOF corrected channel data, while the Clinical Task Network classifies these images. The network integrates feedback from the Clinical Task Network to refine the beamforming process, ensuring that the generated images are optimized for the clinical task.}
    \label{fig:JBC}
\end{figure*}

\begin{figure*}[ht]
    \centering
    \includegraphics[width=0.8\textwidth]{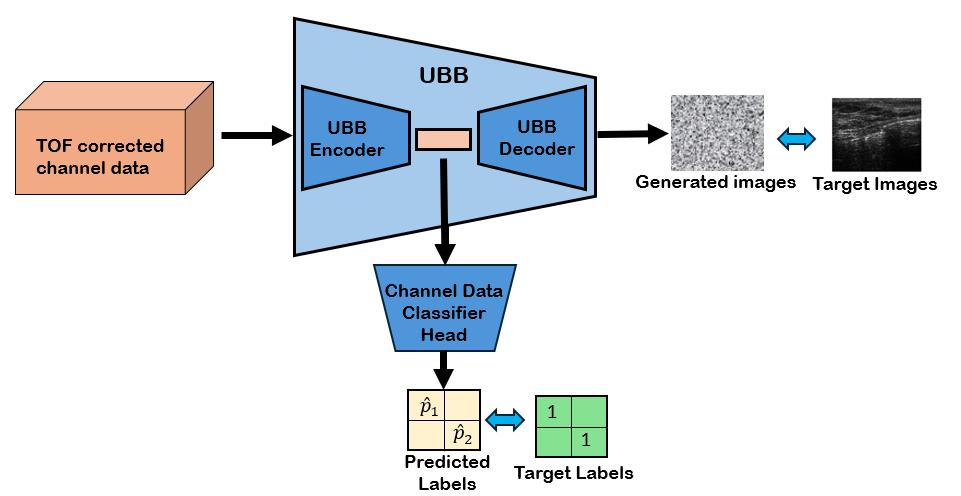}
    \caption{In the CDCB approach, the UNet model is extended with a classification head at the bottleneck, allowing the network to perform both beamforming and classification within a single model. The UNet generates beamformed images, while the classification head processes and classifies the channel data representation in the bottleneck, optimizing both tasks simultaneously and enhancing the clinical relevance of the images.}
    \label{fig:CDCB}
\end{figure*}

We use a customized UNet-based architecture as our baseline DL-based beamformer. This model performs the beamforming process end-to-end, taking time-of-flight (ToF) corrected ultrasound channel data as input and generating US images as output. We refer to this baseline model as the UNet-based Beamformer (UBB). We explore two approaches to integrate clinical task information into the baseline model. In the first approach, we adopt a multi-task strategy that involves training both the UBB model and a ViT image classifier simultaneously. The approach is designed to optimize the beamforming process while simultaneously improving image classification performance. By adopting a multitask strategy, a relationship is established between beamforming and image classification, allowing enhancements in one area to directly influence the other. This task-based beamformer is referred to as the Joint Beamformer and Classifier (JBC) (see Fig. \ref{fig:JBC}). In the second approach, channel data classification is integrated directly into the beamforming process. A classification head is added at the UNet's bottleneck, enabling the model to learn both beamforming and channel data classification concurrently. Classifying the ultrasound channel data at the bottleneck serves as a regularizer to the beamforming process, enhancing clinically relevant features and making them more visible in the final generated image. This integrated beamformer is called the Channel Data Classifier Beamformer (CDCB) (see Fig. \ref{fig:CDCB}).

\subsection{Network architecture}
\label{sec:Architecture}

\subsubsection{DoubleConv Block}
\label{sec:doubleconv}

The DoubleConv block is the basic building block of our baseline DL-based beamformer.
This block comprises two successive sub-blocks: a 3x3 Convolution, Batch Normalization (BatchNorm), and an activation function—either Rectified Linear Unit (ReLU) or Parametric Rectified Linear Unit (PReLU). These sub-blocks are abbreviated as Conv-BatchNorm-PReLU or Conv-BatchNorm-ReLU. Specifically, we employ PReLU activation when this block is applied within the UNet's encoder and ReLU when used in the decoder.

To enhance the stability of our architecture and address the issue of vanishing gradients, the DoubleConv block incorporates a skip connection linking the input and the activation layer of the first sub-block.

\subsubsection{UNet-based Beamformer}
\label{sec:doubleconv}

Our UNet architecture is structured such that the encoder begins with a sequence of three down-scaling blocks implemented using the \textit{Down} module, which combines max-pooling and double convolution (DoubleConv). Following these down-scaling blocks, a bottleneck layer is added, consisting of a DoubleConv block to refine the features extracted by the encoder. In the decoder, the tensor is sequentially processed through three up-scaling blocks implemented using the \textit{Up} module. Each \textit{Up} module applies upsampling followed by double convolution. After the up-scaling blocks, the output tensor is resized to 1024x1024 pixels using bilinear interpolation. Two additional DoubleConv blocks further refine the output features before the final output layer. The final output layer uses a 2D convolution (OutConv) followed by a sigmoid activation function to generate the beamformed ultrasound image. Each convolutional layer in the encoder and decoder is followed by batch normalization to stabilize training and improve convergence. Additionally, dropout layers with a probability of $p=0.5$ are included in the convolutional blocks to enhance robustness and prevent overfitting. Activation functions are selected based on their position in the network. The encoder uses Parametric ReLU (PReLU) activations to accommodate variations in input data, while the decoder uses standard ReLU activations to enhance non-linear transformations. This architecture effectively balances feature extraction, resolution recovery, and output refinement, enabling high-quality ultrasound image reconstruction through end-to-end learning.  

\subsubsection{Task-based Beamformers - Image Classification}
\label{sec:tbic}

The JBC task-based beamformer integrates a UNet-based beamformer with a ViT-based image classifier. The ViT consists of a Patch Embedding layer, a Transformer Encoder with four encoder blocks, and a classification head. The Patch Embedding layer converts 32×32 patches into 512-dimensional embeddings. Each Transformer Encoder block comprises Layer Normalization (LayerNorm), Multi-Head Attention (MHA) with four heads, another LayerNorm, and a Feed Forward (FF) block with expansion factor 4. Skip connections link the input to the MHA and FF outputs. The classification head consists of a LayerNorm followed by a Fully Connected (FC) layer, mapping embeddings to two output classes: Benign and Malignant, for classifying breast lesions.

\subsubsection{Task-based Beamformers - Channel Data Classification}

The CDCB beamformer extends the UNet-based model by incorporating a classification head at the bottleneck, jointly optimizing beamforming and classification. The input is a 128-channel data tensor, where channels represent transmitting elements. As data progresses through the encoder, the number of channels increases, capturing increasingly abstract and complex features. At the bottleneck, the classification head applies a \textit{ChannelAttention} module to emphasize important channels. The \textit{ChannelAttention} module uses global average pooling along spatial dimensions to extract a 1D vector, followed by two fully connected layers with Leaky ReLU activation. The weights are interpolated across channels to enhance smooth transitions, reducing noise sensitivity. The weighted bottleneck representation is processed through convolutional layers with batch normalization, Leaky ReLU, and average pooling. The first block has 32 filters (3×3), the second has 64. The final fully connected layer outputs classification predictions. The CDCB model produces two outputs: a beamformed ultrasound image and classification predictions. By integrating classification at the bottleneck, it enhances image quality and clinical relevance, refining diagnostic features while optimizing beamforming.

\label{sec:cdcb}

\subsection{Training Strategy}
\label{sec:strategy}
During the training of our DL-based beamformers, we used US channel data as input and MV images as target outputs.
The shared objective among all our deep learning-based beamformers is image generation, and the conventional loss function used for such tasks is the Mean Squared Error (MSE) loss.
However,  MSE loss has a limitation, as it tends to converge toward the mean of plausible solutions during training. This can lead to the generation of images with diminished resolution.
To further improve our loss objective, we regularize our models in an additional way, which does not focus on the pixel values but focuses on maintaining the general luminance, contrast, and structure of the scanned object in the image. For that, we used the Multi-Scale Structural Similarity (MS-SSIM) loss in addition to the MSE loss.
Structural Similarity (SSIM) is a measure for comparing two signals $x,y$ in terms of luminance, contrast and structure. The metrics are given by:
\begin{equation}
\begin{aligned}
     l\left( x, y \right)  =  \frac{2\mu_x\mu_y+C_1}{\mu_x^2+\mu_y^2+C_1}
\end{aligned}
\label{eq:luminance}
\end{equation}
\begin{equation}
\begin{aligned}
     c\left( x, y \right)  =  \frac{2\sigma_x\sigma_y+C_2}{\sigma_x^2+\sigma_y^2+C_2}
\end{aligned}
\label{eq:contrast}
\end{equation}
\begin{equation}
\begin{aligned}
     s\left( x, y \right)  =  \frac{2\sigma_{xy}+C_3}{\sigma_x\sigma_y+C_3}
\end{aligned}
\label{eq:structure}
\end{equation}
where, $\mu_{x}, \mu_{y}, \sigma_{x}, \sigma_{y} $ and $\sigma_{xy}$ are the means, standard deviations and cross-correlation of the two images, calculated over an 11-pixel 1D sliding Gaussian window (with $\sigma=1.5$) 
and $C_{1}$, $C_{2}$ are constants meant to stabilize the division in image regions where the local means or standard deviations are close to zero. We choose the stabilization constants to be $C_{1}=(k_1 \cdot L)^2$,   $C_{2}=(k_2 \cdot L)^2$ where $L$ is the dynamic range of pixel values and $k_{1}=0.01, k_{2}=0.03$ as suggested by Wang \emph{et al.} \cite{Wang2004}.

The structural similarity is defined by: 
\begin{equation}
\begin{aligned}
     SSIM\left( x, y \right)  =  [l(x,y)]^\alpha\cdot[c(x,y)]^\beta\cdot[s(x,y)]^\gamma
\end{aligned}
\label{eq:ssim}
\end{equation}
where $\alpha,\beta,\gamma$ are parameters that control the weight of each component.
The MS-SSIM metric is computed by performing a series of steps that involve iteratively applying a low-pass filter and down-sampling to the two images being compared. For each scale in this process, we calculate the contrast and structure components as previously defined, while the luminance component is determined exclusively at the highest scale. The final MS-SSIM score is derived by combining the measurements from all scales using the following expression:

\begin{equation}
\begin{multlined}
MS-SSIM\left( x, y \right) =\\
\left[l_M(x,y)\right]^{\alpha_{M}} \prod_{m=1}^M\left[c_m(x,y)\right]^{\beta_{m}}\left[s_m(x,y)\right]^{\gamma_{m}}. \
\end{multlined}
\label{eq:MS-SSIM}
\end{equation}

Where $M$ is the number of scales and $\alpha_{M},\beta_{m},\gamma_{m}$ are parameters that control the weight of each component at each scale $m$.
In our loss the number of scales we use is $M=5$ and $\beta_{1}=\gamma_{1}=0.0448$, $\beta_{2}=\gamma_{2}=0.2856$, $\beta_{3}=\gamma_{3}=0.3001$, $\beta_{4}=\gamma_{4}=0.2363$, $\alpha_{5}=\beta_{5}=\gamma_{5}=0.1333$.

The MS-SSIM is bounded by $1$ for all $x,y$ where
\begin{align*}
MS-SSIM(x,y)=1 \leftrightarrow x=y     
\end{align*}
Therefore we define the MS-SSIM loss to be:

\begin{equation}
\begin{aligned}
     \mathcal{L}_{MS-SSIM}\left( \hat{y}, y \right)  =  \left.1-MS-SSIM(\hat{y},y)\right.
\end{aligned}
\label{eq:MS-SSIM-Loss}
\end{equation}
where $\hat{y}$ is the generated image, and $y$ is the target MV image. Finally, the loss of our baseline UNet-based beamformer is defined by:
\begin{equation}
\begin{multlined}
     \mathcal{L}_{UBB}\left( \hat{y}, y \right) =
     \frac{\lambda}{N}.\|\hat{y}-y\|_{2}^2 + \\
     (1-\lambda)\cdot\mathcal{L}_{MS-SSIM}(\hat{y},y)
\end{multlined}
\label{eq:UB-Loss}
\end{equation}
where $N$ represents the total number of pixels in $y$, and $\lambda$ acts as a parameter to balance the weight between the MSE loss and the MS-SSIM loss. We have set $\lambda$ to 0.8. The baseline UNet-based beamformer was trained for 20 epochs with a learning rate of $10^{-4}$ using the Adam optimizer with $\beta_1=0.5$, $\beta_2=0.999$.

The training of the JBC model is conducted over 10 epochs, divided into two phases. In the first phase, spanning the initial 5 epochs, the UBB and the ViT are trained simultaneously. During this phase, the UBB is trained using the \(\mathcal{L}_{UBB}\) loss, while the ViT is trained on the target MV images using cross-entropy loss. In the second phase, which lasts for the remaining 5 epochs, classification feedback is introduced into the training process. The ViT is trained using the feedback loss:

\begin{equation}
\begin{multlined}
    \mathcal{L}_{\text{feedback}}(\hat{y}, y, \ell) =  \\ \mathcal{L}_{CE}\left(\mathcal{V}_{\text{classifier}}(\hat{y}), \ell\right) + \mathcal{L}_{CE}\left(\mathcal{V}_{\text{classifier}}(y), \ell\right).
\end{multlined}
\label{eq:Feedback-Loss}
\end{equation}

Where \(\mathcal{L}_{CE}\) is the cross-entropy loss function, \(\mathcal{V}_{\text{classifier}}(\hat{y})\) and \(\mathcal{V}_{\text{classifier}}(y)\) represent the classifier outputs for the generated image \(\hat{y}\) and the target MV image \(y\), respectively, and \(\ell\) is the true label. During this second phase, the UBB is trained with the overall JBC loss, which combines the beamforming loss with the classifier feedback loss:

\begin{equation}
\begin{aligned}
    \mathcal{L}_{JBC}(\hat{y}, y, \ell) = \gamma\cdot\mathcal{L}_{UBB}\left(\hat{y}, y\right) + \mathcal{L}_{\text{feedback}}\left(\hat{y}, y, \ell\right).
\end{aligned}
\label{eq:JBC-Loss}
\end{equation}

Where \(\gamma\) is a scaling factor that adjusts the balance between the beamforming loss and the classification feedback loss. In our experiments, we set \(\gamma = 100\).

Each network, the UBB and the ViT, is trained using its own Adam optimizer. Both optimizers are configured with a learning rate of \(10^{-4}\), \(\beta_1 = 0.5\), and \(\beta_2 = 0.999\).

This two-phase training strategy allows the model to initially focus on optimizing the beamforming and classification tasks separately before integrating the feedback mechanism. In the final phase, the joint optimization of both tasks ensures that the generated US images align more closely with the clinical objectives.

The training of the CDCB model involves a two-phase multitask approach, similar to the training strategy employed for the JBC model, enabling both the beamforming and classification tasks to be optimized effectively. The model, which features a classification head integrated at the UNet bottleneck, is trained over a total of 15 epochs using a combined loss function. In the first phase, spanning the initial 5 epochs, the network is trained solely for the beamforming task, without updating the classification head. This phase focuses on optimizing the beamforming objective. During the second phase, which lasts for the remaining 10 epochs, the classification head is included, and both beamforming and classification tasks are trained simultaneously. The combined loss function used during this phase is defined as:

\begin{equation}
\begin{aligned}
    \mathcal{L}_{\text{CDCB}}(\hat{y}, y, \hat{\ell}, \ell) = \mathcal{L}_{UBB}\left(\hat{y}, y\right) + \mathcal{L}_{CE}\left(\hat{\ell}, \ell\right)
\end{aligned}
\label{eq:CDCB-Loss}
\end{equation}

where \(\mathcal{L}_{UBB}(\hat{y}, y)\) is the beamforming loss calculated between the generated image \(\hat{y}\) and the target MV image \(y\), and \(\mathcal{L}_{CE}(\hat{\ell}, \ell)\) is the cross-entropy loss calculated between the predicted labels \(\hat{\ell}\) and the true labels \(\ell\). The model is trained using the Adam optimizer, with a learning rate of \(10^{-4}\), \(\beta_1 = 0.5\), and \(\beta_2 = 0.999\). The simultaneous learning of beamforming and classification in this approach allows for mutual feedback between the two tasks, enhancing the overall performance of the model.
All models were trained using a single NVIDIA A40 GPU with 48GB of memory.

\section{Channel Data Augmentations}
\label{sec:cd_aug}
In this study, we applied a carefully selected set of augmentations to the input channel data, drawing inspiration from self-supervised models commonly employed for solving inverse problems. These models often involve training on corrupted versions of the data while ensuring the integrity of the target is preserved. Adopting this principle, we introduced controlled perturbations to the channel data through targeted augmentation techniques designed to enhance model robustness. These augmentations simulate real-world ultrasound corruptions while incorporating non-realistic distortions to reduce overfitting. However, experimenting with these augmentations revealed that they can sometimes be overly powerful, potentially smoothing out critical clinical features of the image. To address this, we carefully balanced the augmentation probabilities and the number of training epochs to preserve clinically relevant details while still benefiting from the regularization effects of the augmentations. This adaptation of self-supervised learning insights to ultrasound beamforming seeks to improve the overall performance and generalization of our DL-based beamformer.

\subsubsection{Noise Augmentations}

We introduce a new noise augmentation technique specifically designed for training DL-based US beamformers: speckle noise augmentation. This new augmentation fits the unique characteristics of the US beamforming scenario, enhancing the model's ability to handle the inherent noise in US imaging. The speckle noise simulation algorithm, adapted from \cite{speckle}, has been tailored to work with US channel data. The original algorithm, designed for adding speckle noise to ultrasound images, was modified to treat the time samples obtained by each transmitting element for all transmission events as an image. This allowed us to apply the speckle noise algorithm to the received data from each element.

   \begin{algorithm}  
		\caption{Add speckle noise to US channel data}
		\label{alg:addSpeckles}
		\KwData{\\
                    X- US channel data tensor of shape $(B,C,N_{s},N_{t})$\\
                    B- Batch size \\
                    C- Number of transmitting elements \\
                    $N_{s}$ - Number of time samples \\
                    $N_{t}$ - Number of transmissions \\
                    $\sigma^2$ - Noise level \\ 
                    $\Tilde{C}$ - Number of transmitting elements for which noise is added \\ 
                    $\mathcal{K}$ - Kernel for weighting the noise. }.
        Slice the channel data: $\Tilde{X}_{i,\Tilde{j},k,l} \gets X_{i,j,k,l}$ \\ \nonl
        $S=\lfloor\frac{C}{\Tilde{C}}\rfloor, \Tilde{j}\in[\Tilde{C}],\: j\in\{p\:\vert \: 
   p\in[C] \wedge p\vert S\}$
        
	 Sample $G_{x},G_{y}\sim\mathcal{N}(\textbf{0},\sigma^2\cdot I_{B\times \Tilde{C}\times N_{s}\times N_{t}})$
  
  Apply 2D convolution: $U=G_{x}\ast\mathcal{K},V=G_{y}\ast\mathcal{K}$

  Calculate element-wise sign function: 
    $\Tilde{X}_{i,\Tilde{j},k,l}^{sgn}\gets sgn(\Tilde{X}_{i,\Tilde{j},k,l})$
  
  Calculate element-wise absolute value: 
    $\Tilde{X}_{i,\Tilde{j},k,l}^{abs}\gets \lvert X_{i,\Tilde{j},k,l}\rvert$

   Add the speckle noise to the sliced tensor: 
  $\Tilde{X} \gets \Tilde{X}+2\cdot \Tilde{X}^{sgn}\circ \Tilde{X}^{abs}\circ U+U^{\circ2}+V^{\circ2}$

  Embed the noisy channels back in the original tensor:
  $X_{i,j,k,l}=\Tilde{X}_{i,\Tilde{j},k,l}$
  
    \KwOut{$X$}
	\end{algorithm}

In our implementation we used noise level $\sigma^2=4.5$, number of noisy channels $\Tilde{C}=25$ and we used the following kernel:
 \begin{equation}
     \mathcal{K}= \frac{1}{2.9}\begin{bmatrix}
0.9 & 0.9 & 0.9 \\
0.8 & 0.8 & 0.8 \\
0.6 & 0 & 0.6 \\
0.4 & 0.4 & 0.4 \\
0.2 & 0.2 & 0.2 
\end{bmatrix}.
\label{eq:kernel}
 \end{equation}

By directly applying the speckle noise algorithm from \cite{speckle} to the entire image, without any sampling or interpolation as demonstrated in \cite{speckle}, one can derive the following equation for the pixel value in the noisy image:

\begin{equation}
\begin{split}
    \Tilde{I}[x,y]=I[x,y]+2\cdot\sqrt{I[x,y]}\cdot\sum_{i=1}^{M}{u_{i}} + \\
 + \left(\sum_{i=1}^{M}{u_{i}}\right)^2+\left(\sum_{j=1}^{M}{v_{j}}\right)^2 .
\end{split}
\label{eq:noisyImage}
\end{equation}

Where $I$ is the original image, $\Tilde{I}$ is the image with speckle noise, $M\sim U(a,b)$ random number of phasors where $a,b$ are the minimum and maximum limits for the number of phasors and $\{u\}_{i=1}^M,\{v\}_{j=1}^M \sim \mathcal{N}(0,\sigma^2)$, where $\sigma^2$ is the noise level. As a point of comparison, after applying our proposed algorithm, one can utilize the same mathematical analysis on a specific channel within our channel data, and get the following relation between a noisy sample to the original sample in the channel data: 

\begin{multline}
 \Tilde{X}_{e}[x,y]= X_{e}[x,y] + \\
 2 \cdot \text{sgn}(X_{e}[x,y]) \cdot \sqrt{\lvert X_{e}[x,y] \rvert} \cdot \sum_{i=1}^{N+1} k_i \cdot u_i + \\
  \left( \sum_{i=1}^{N+1} k_i \cdot u_i \right)^2 + \left( \sum_{j=1}^{N+1} k_j \cdot v_j \right)^2.
\end{multline}

Here, \( X_e \) represents the samples in the channel data tensor corresponding to element \( e \), while \( \Tilde{X}_e \) denotes the same samples with added speckle noise. The parameter \( N \), the number of phasors, is fixed at 14 in our case, unlike in \cite{speckle} where it is randomly chosen. The variables \( \{u\}_{i=1}^{N+1}, \{v\}_{j=1}^{N+1} \) are sampled from the normal distribution \( \mathcal{N}(0, \sigma^2) \), where \( \sigma^2 \) represents the noise level. The constants \( k_i \) and \( k_j \) are defined as \( k_i = k_j = \mathcal{K}_{mt} \), where \( \mathcal{K} \) is the kernel provided in Eq. \eqref{eq:kernel}, and \( m = \lceil \frac{i}{3} \rceil \) and \( t \equiv (i-1) \pmod{3} + 1 \).

We used only tensor operations and obtained a relation between the noisy and the original channel data which is similar to \eqref{eq:noisyImage}. Therefore, we managed to generalize the algorithm in \cite{speckle} efficiently for batches of channel data tensors.
The speckle noise is added to the input channel data with probability $p=0.5$.
In addition to speckle noise, we also applied standard Gaussian noise augmentations, both additive and multiplicative. When noise was introduced with a probability of $p=\frac{1}{3}$, we randomly selected between additive Gaussian noise $\mathcal{N}(0,50)$ and multiplicative Gaussian noise $\mathcal{N}(1,0.8)$, with each type of noise having an equal likelihood of being chosen. Gaussian noise augmentations are well-known and widely used not only for ultrasound beamforming models but also across various other domains. Like the speckle noise augmentation, the goal of these augmentations is to make the model more robust by introducing variability into the training data.
\\

\subsubsection{Channel Data SpecAugment}

In this work, we introduce a new data augmentation technique called \textit{Channel Data SpecAugment}, specifically designed for ultrasound channel data. This method is inspired by the original SpecAugment augmentation technique proposed by Park et al. \cite{Park_2019}, which was primarily developed for augmenting speech data. The original SpecAugment method first converts the input speech signal into a spectrogram — a time-frequency representation of the data. The spectrogram is then augmented by stretching or compressing it along the time axis. Afterward, a contiguous range along the time axis is randomly selected and masked, followed by masking a different contiguous range along the frequency axis. These operations increase data variability and help models generalize better.

Our \textit{Channel Data SpecAugment} modifies and extends this approach to better suit the characteristics of ultrasound channel data. In addition to stretching or compressing the spectrogram in the time domain, we also apply similar operations in the spatial domain. This ensures that the model becomes robust to distortions in both the temporal and spatial dimensions. Furthermore, we apply random spatial masking alongside the random time and frequency masking used in the original SpecAugment. By randomly masking sequences of spectrograms in the spatial domain, our method enhances the model's ability to handle potential loss or corruption of spatial information.

\begin{algorithm}
\caption{Channel Data SpecAugment}
\label{alg:specaugment}
\KwData{\\
    $X \in \mathbb{R}^{C \times N_{t} \times N_{s}}$ - Ultrasound channel data tensor \\
    $C$ - Number of transmitting elements \\
    $N_{t}$ - Number of time samples \\
    $N_{s}$ - Number of transmission events \\
    $M_{t}, M_{f}, M_{s}$ - Maximum mask lengths for time, frequency, and space dimensions
}

\begin{enumerate}
  \item Rearrange $X$ to shape $(C, N_{s}, N_{t})$, then calculate the spectrogram: $S = \text{Spectrogram}(X) \in \mathbb{R}^{C \times N_{s} \times N_{f} \times N_{\tau}}$
  
  \item Spectrogram stretching:
  \begin{itemize}
    \item Sample $b \sim \text{Bernoulli}\left(\frac{1}{2}\right)$
    \item \textbf{If} $b = 1$: Stretch spectrograms in time dimension with $\alpha \sim \mathcal{U}(0.9, 1.2)$.
    \item \textbf{Else}: Stretch in transmission events dimension with $\beta \sim \mathcal{U}(0.95, 1.1)$.
  \end{itemize}
  
  \item Masking:
  \begin{itemize}
    \item For $d \in \{N_{\tau}, N_{f}, N_{s}\}$, sample $m_d$ from the discrete set $\{0,1,...,M_d\}$
    \item Mask a randomly selected contiguous range of length $m_d$ along the corresponding dimension $d$ across all spectrograms.
  \end{itemize}
  
  \item Calculate the inverse spectrogram using least squares estimation: $X' = \text{InverseSpectrogram}(S')$
  
  \item Adjust $X'$ to match the dimensions of $X$ by slicing or zero padding.
  
  \item Return $X'$
\end{enumerate}

\end{algorithm}

\subsubsection{Channel Data Subsampling and Masking}
Inspired by \cite{mamistvalov2022deep}, which proposed a deep-learning-based reconstruction of B-mode images using temporally and spatially sub-sampled channel data, we adopt their subsampling techniques as a data augmentation strategy. In \cite{mamistvalov2022deep}, subsampling was applied in the time domain and the transmitting elements domain. To further enhance robustness against spatial corruption, we extend this approach by introducing subsampling in the transmission events domain. Specifically, we subsample the time and transmission events dimensions by a factor of 2. In the transmission elements domain, unlike the fixed subsampling used in [1], we randomly select between 25\% and 50\% of the elements to include in the final augmented channel data tensor, masking all other elements' channels with zeros. This combined strategy strengthens the model's ability to manage variations and potential data corruption across different dimensions.

Moreover, to prevent overfitting in our deep learning-based beamformer, we adapt an augmentation technique known as Coarse Dropout to fit the case of ultrasound channel data. In this adaptation, we treat the time and transmission events dimensions as an image and the transmitting elements dimension as channels. With a probability of \(p = \frac{1}{3}\), we randomly sample 5 patches of size \(64 \times 16\) to mask across all channels within the channel data tensor. This encourages the model to identify robust features even when information is partially obscured.

\section{Results}
\label{sec:res}

We compared our proposed task-based US beamformers against conventional techniques, namely DAS and MV. Our evaluation employs a test set of in-vivo images, analyzing performance through both qualitative observations and quantitative analysis using contrast metrics. The study also investigates the impact of incorporating channel data augmentations into the training process, revealing an overall enhancement in beamforming efficacy. Notably, the task-based beamformers demonstrated superior performance across all evaluation metrics, outperforming the traditional DAS and MV methods. This advancement underscores the potential of task-based approaches in refining US imaging quality through both innovative algorithmic enhancements and data augmentation strategies. 

\begin{figure*}[!t]
  \centering
  \begin{subfigure}[b]{\textwidth}
    \centering
    \includegraphics[width=\textwidth]{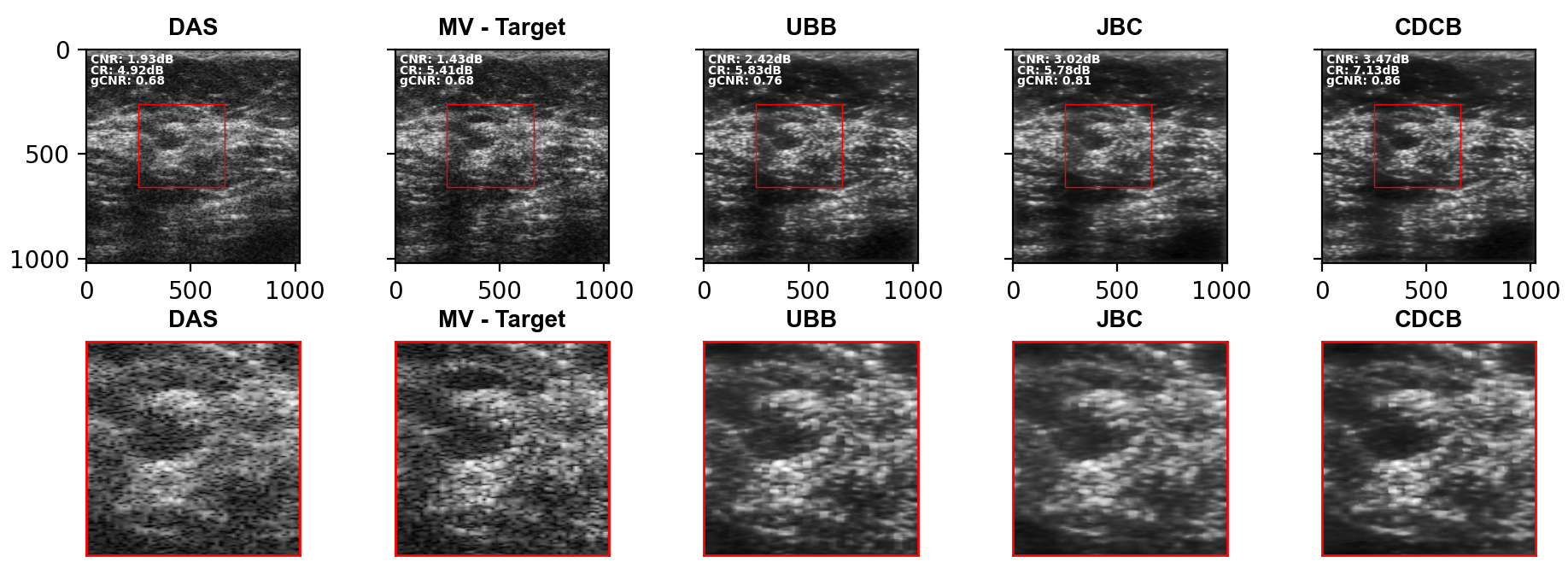}
    \caption{Benign lesion 1. The image shows the original view (top) and a zoomed-in view of the ROI (bottom).}
  \end{subfigure}

  \vspace{1em} 
  
  \begin{subfigure}[b]{\textwidth}
    \centering
    \includegraphics[width=\textwidth]{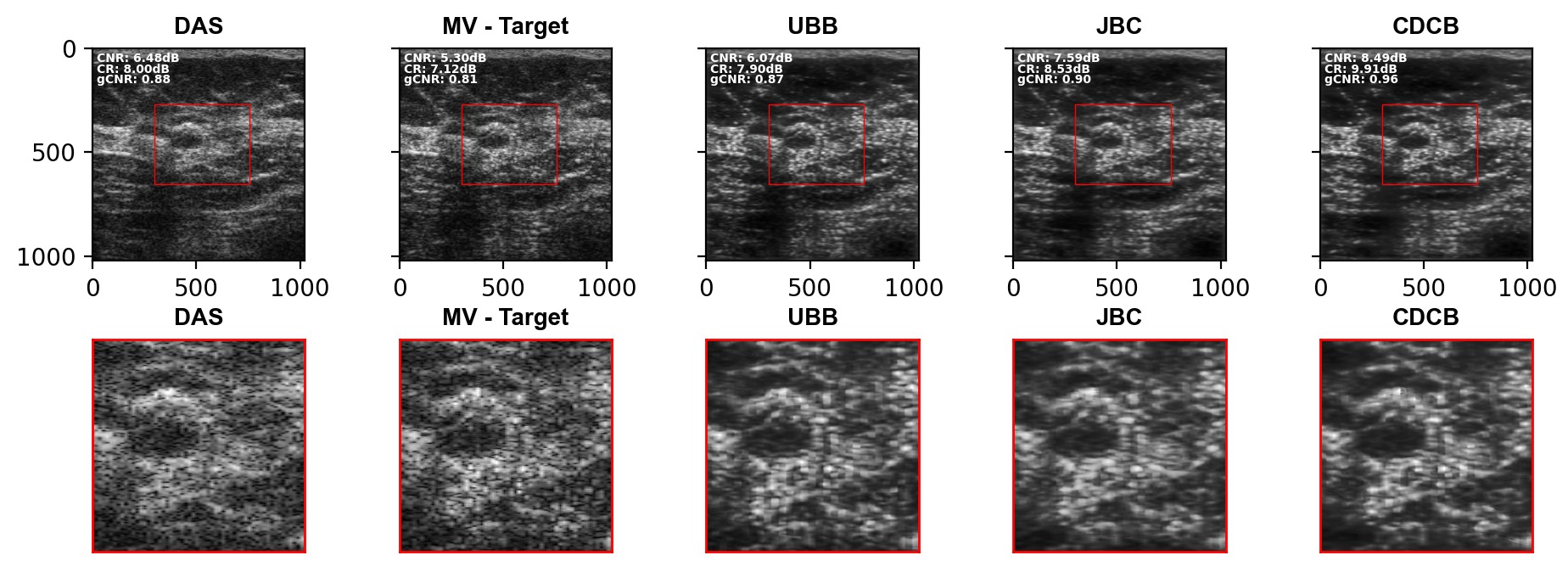}
    \caption{Benign lesion 2. The image shows the original view (top) and a zoomed-in view of the ROI (bottom).}
  \end{subfigure}

  \caption{Comparison of beamforming techniques on images of benign lesions. Each subfigure depicts a different benign lesion, showing the original image (top) and the zoomed-in view of the ROI (bottom).}
  \label{fig:benign_comparison}
\end{figure*}

\begin{figure*}[!t]
  \centering
  \begin{subfigure}[b]{\textwidth}
    \centering
    \includegraphics[width=\textwidth]{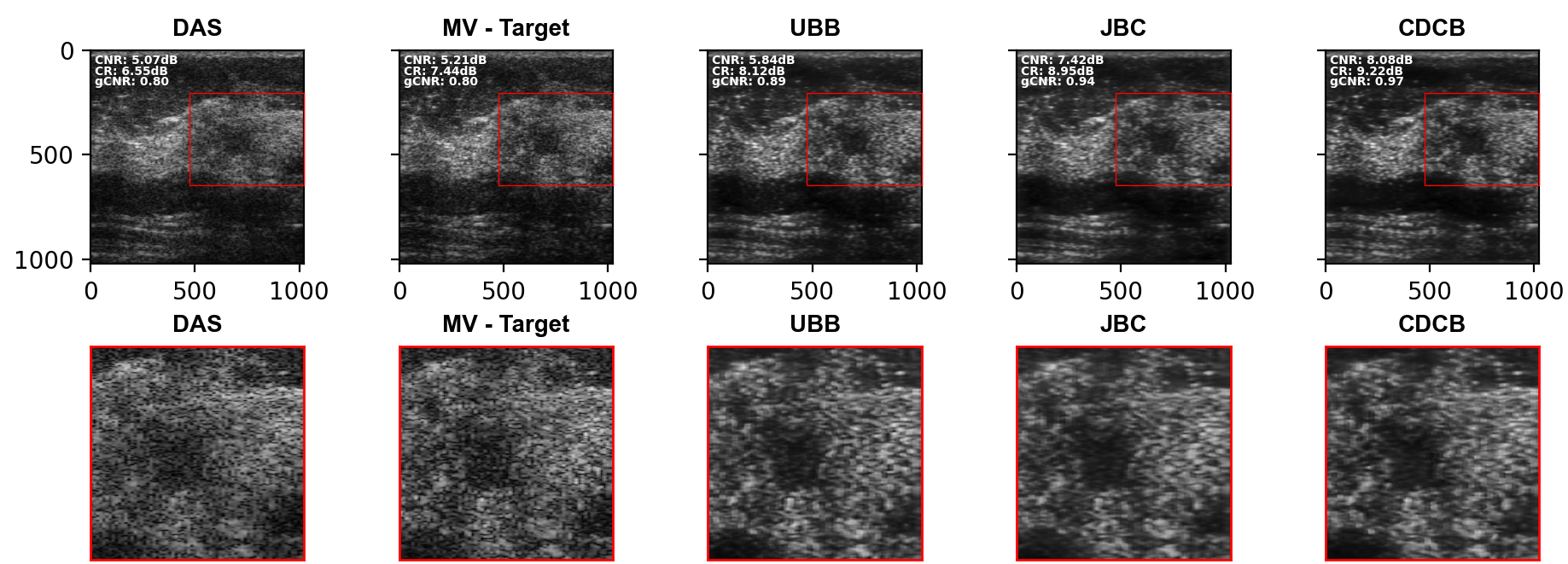}
    \caption{Malignant lesion 1. The image shows the original view (top) and a zoomed-in view of the ROI (bottom).}
  \end{subfigure}

  \vspace{1em} 
  
  \begin{subfigure}[b]{\textwidth}
    \centering
    \includegraphics[width=\textwidth]{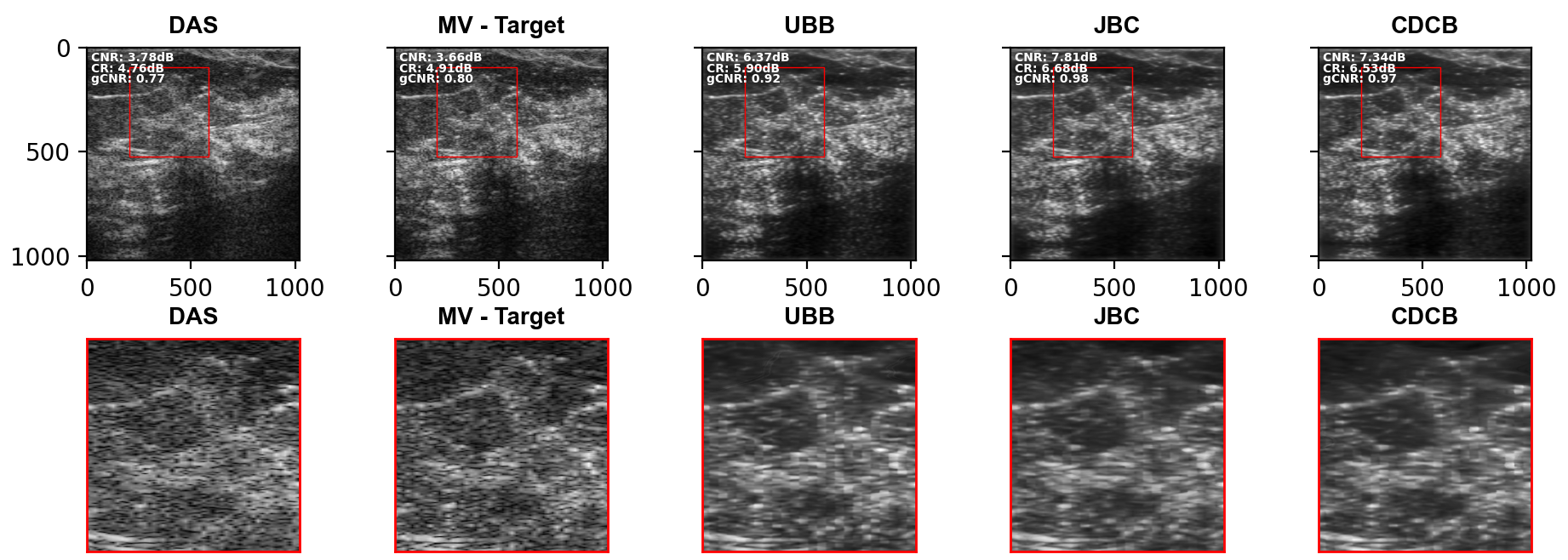}
    \caption{Malignant lesion 2. The image shows the original view (top) and a zoomed-in view of the ROI (bottom).}
  \end{subfigure}

  \caption{Comparison of beamforming techniques on images of malignant lesions. Each subfigure depicts a different malignant lesion, showing the original image (top) and the zoomed-in view of the ROI (bottom).}
  \label{fig:suspicious_comparison}
\end{figure*}

\subsection{Experimental Setup}
\label{sec:expSetup}
For training and testing our model, in-vivo breast US channel data was acquired from 20 patients by a Vantage system (Verasonics Inc., WA, USA) with 128 elements L11-5v probe using a line by line focus beam acquisition scheme. The central frequency of the scans was 7.6MHz and the sampling rate was 31.25MHz, resulting in 1579 samples per image line, that represent depth of 39mm. Approximately 1500 frames were captured for each patient in which the lesion was visible. Out of the 20 patients, malignant breast masses were present in 4 patients, while the rest were present with benign lesions. Those benign lesions were either fibroadenoma, a benign solid mass, or cyst which is a benign fluid filled lesion. All breast US scans were conducted at the department of radiology, Rabin Medical Center, under Helsinki committee approval number 0085-19-RMC. An offline ToF correction was applied to the channel data from all sources using Matlab 2020a similarly to \cite{mamistvalov2021deep}. The resulting ToF corrected channel data, a 3 dimensional tensor of size 128x1579x128 (channels, axial depth and lateral scan lines dimensions), was used as an input to our model. In addition, offline DAS and MV beamformers were applied, using Matlab 2020a, to the channel data in order to obtain B-mode images in full clinical size (1024x1024 pixels). Full-size MV B-mode images were used as a training target while DAS images were used as an additional reference for comparison.

\subsection{Metrics for evaluation}
\label{section:metrics}
We used CNR, gCNR, and CR to evaluate the beamformed ultrasound images after histogram matching to DAS images. These contrast metrics are calculated between a region of interest within the lesion and the surrounding tissue, and are defined as follows:

CNR is calculated using the following equation:
\begin{equation}
  CNR = 20\cdot\log_{10}\left(\frac{\mid \mu_{c}-\mu_{b} \mid}{\sqrt{\sigma_{c}^2 + \sigma_{b}^2}}\right)
\end{equation}
where $\mu_{c}$ and $\mu_{b}$ are the means, and $\sigma_{c}$ and $\sigma_{b}$ are the standard deviations of the gray scale values within the lesion (target) and the background tissue, respectively.

gCNR is less sensitive to large dynamic ranges and is calculated using:
\begin{equation}
  gCNR = 1-\int_{-\infty}^\infty \min_{x}\{{p_c(x),p_b(x)}\}dx,
\end{equation}
where \(p_c\) and \(p_b\) are the probability density functions of the grayscale values inside the lesion and in the background, respectively.

CR is a measure of the contrast between the lesion and the surrounding tissue and is given by:
\begin{equation}
  CR = -20\cdot\log_{10}\left(\frac{\mu_{\text{lesion}}}{\mu_{\text{background}}}\right)
\end{equation}
where $\mu_{\text{lesion}}$ and $\mu_{\text{background}}$ are the mean gray scale values within the lesion and the background tissue, respectively. CR quantifies the degree of contrast between the lesion and the surrounding tissue, with higher values indicating better contrast.

\begin{table}[ht]
\centering
\caption{Comparison of contrast metrics (gCNR, CNR, CR) between lesions and surrounding tissue for each DL-based beamformer trained with and without channel data augmentations.}
\label{table:augmentations_results}
\begin{tabular}{|c|c|c|c|}
\hline
Method                 & Metric   & w/o CD augmentations  & w/ CD augmentations           \\ \hline
\multirow{3}{*}{UBB}& CNR (dB) & $5.23\,(\pm3.98)$               & $5.23\,(\pm3.56)$               \\ \cline{2-4} 
                       & gCNR     & $0.83\,(\pm0.15)$               & $0.85\,(\pm0.14)$               \\ \cline{2-4} 
                       & CR (dB)  & $8.65\,(\pm3.44)$              & $8.60\,(\pm3.46)$              \\ \hline
\multirow{3}{*}{CDCB}& CNR (dB) & $5.46\,(\pm2.94)$               & $5.65\,(\pm3.59)$               \\ \cline{2-4} 
                       & gCNR     & $0.86\,(\pm0.14)$               & $0.88\,(\pm0.14)$               \\ \cline{2-4} 
                       & CR (dB)  & $8.40\,(\pm3.45)$              & $8.91\,(\pm3.13)$              \\ \hline
\multirow{3}{*}{JBC}& CNR (dB) & $5.02\,(\pm3.47)$               & $5.25\,(\pm3.84)$               \\ \cline{2-4} 
                       & gCNR     & $0.83\,(\pm0.15)$               & $0.86\,(\pm0.14)$               \\ \cline{2-4} 
                       & CR (dB)  & $8.58\,(\pm3.67)$              & $8.31\,(\pm3.26)$              \\ \hline
\end{tabular}
\end{table}

\begin{table}[ht]
\centering
\caption{Comparison of contrast metrics (gCNR, CNR, CR) between lesions and surrounding tissue. The table compares traditional beamformers (DAS and MV) with a basic DL-based beamformer and task-based beamformers.}
\label{table:beamformers_comparison}
\small
\begin{tabular}{@{}>{\centering\arraybackslash}m{0.3\linewidth}ccc@{}}
\toprule
\textbf{Method} & \textbf{CNR (dB)} & \textbf{CR (dB)} & \textbf{gCNR} \\ \midrule
DAS & $3.40\,(\pm3.57)$ & $6.33\,(\pm1.97)$ & $0.73\,(\pm0.16)$ \\
MV & $3.07\,(\pm4.12)$ & $6.37\,(\pm2.08)$ & $0.72\,(\pm0.16)$ \\
UBB + CD AUG & $5.23\,(\pm3.56)$ & $8.60\,(\pm3.46)$ & $0.85\,(\pm0.14)$ \\
CDCB + CD AUG & $5.65\,(\pm3.59)$ & $8.91\,(\pm3.13)$ & $0.88\,(\pm0.14)$ \\
JBC + CD AUG & $5.25\,(\pm3.84)$ & $8.31\,(\pm3.26)$ & $0.86\,(\pm0.14)$ \\
\bottomrule
\end{tabular}
\end{table}

\subsection{Numerical Results}
\label{sec:numerical_results}
The test set used for model comparison is derived from 4 patients, 2 with benign lesions and 2 with malignant lesions. Each patient contributed 100 frames for analysis. Accompanying each frame is a mask that highlights a region of interest within the lesion and its surrounding background.  The masks are subsequently used to calculate the contrast metrics. Initially, we assessed the performance of the deep learning-based US beamformers, particularly focusing on the impact of incorporating channel data augmentations during their training phase. As evidenced in Table \ref{table:augmentations_results}, the integration of channel data augmentations led to a notable improvement in performance, as measured by contrast metrics. This enhancement underscores the effectiveness of data augmentation in refining the capabilities of US beamformers to yield higher contrast images, thereby facilitating more accurate diagnostics. Subsequent comparisons revealed several key findings. First, as presented in Table \ref{table:beamformers_comparison}, all DL-based beamformers trained with channel data augmentations outperformed the traditional beamforming techniques, DAS and MV, demonstrating the superiority of deep learning approaches in generating higher contrast images. Second, within the category of DL-based methods, all task-based beamformers performed better than the UBB baseline model, which proves that integrating a clinical task into the beamforming process significantly improves performance. Lastly, among the task-based beamformers, the CDCB model emerged as the best performer. This finding suggests that by regulating the channel data representation, rather than the generated US images, one can achieve better overall performance, leading to more clinically relevant ultrasound images.

\subsection{Beamforming Results}
Figures \ref{fig:benign_comparison} and \ref{fig:suspicious_comparison} provide a qualitative comparison of ultrasound images produced by traditional beamforming techniques, DL-based beamformers, and task-based beamformers. The red bounding boxes highlight the region of interest (ROI) around the lesion and the surrounding tissue. The top row of each figure shows the original images with the red bounding box, while the bottom row presents a zoomed-in view of the ROI to assess clarity and resolution. Both DAS and MV techniques show significant noise, making it harder to distinguish the lesion from the surrounding tissue. In contrast, all DL-based beamformers, trained with channel data augmentations, produce clearer and cleaner images where the lesion is more easily detectable and separable from surrounding tissues. For benign lesions, there is improved visibility of posterior acoustic enhancement, aiding in identifying lesion characteristics. For malignant lesions, there is a significant reduction in speckle noise, further improving clarity and aiding in visualization. While the JBC model introduces some speckle reduction compared to UBB and traditional models, the CDCB model outperforms them all, showing the most notable improvements. Overall, the image resolution is improved with the DL-based models, with the task-based beamformers showing slightly better resolution, further enhancing lesion visibility and contrast. This demonstrates that our methods are effective in improving the clinical relevance of ultrasound beamforming, particularly when tailored for a specific clinical task, such as lesion detection.

\section{Conclusion}
\label{sec:conc}

This paper presented a DL-based framework for task-based US beamforming, aimed at enhancing clinical outcomes by integrating clinical tasks directly into the beamforming process. The proposed approaches incorporate task-specific information, such as lesion classification, to optimize both image quality and clinical relevance. Additionally, we introduced channel data augmentations to address the challenges posed by noisy and limited in-vivo data, enhancing the robustness of the beamformers. These augmentations, which included adding noise, subsampling, and masking, simulate real-world data corruptions and were designed to improve the model's ability to handle noisy inputs. Our experimental evaluations demonstrated that training the beamformers with these channel data augmentations resulted in better performance compared to training without augmentations. The models trained with augmentations achieved higher contrast between the lesion and surrounding tissue in the ROI, as measured by image contrast metrics. In the second part of our evaluation, the task-based beamformers, which incorporated clinical task feedback, demonstrated superior performance compared to the baseline beamformer trained without task-specific feedback. Among the task-based beamformers, the CDCB model emerged as the best performer, providing the most significant improvements in image clarity and contrast. In conclusion, the proposed task-based beamforming framework offers a promising approach for improving ultrasound imaging by incorporating clinical tasks into the beamforming process. While our methods have shown potential to enhance diagnostic accuracy and image quality, future work will explore extending this framework to additional clinical tasks. We also aim to incorporate further data augmentations to improve the model's robustness in real-world scenarios.

\bibliographystyle{myIEEEtran}
\bibliography{main}
\end{document}